\documentclass[12pt]{article}
\usepackage{amsmath,amsfonts,amssymb,dcolumn,lscape,subfigure,alltt,authblk}
\usepackage{graphicx,color,soul}
\usepackage{tabularx}

\begin{document}
% Zeilenabstand (doppelt=8.2mm , ca. einfach:5.5mm)
\baselineskip=6.6mm
\bibliographystyle{aip}
\newcommand{\be}{\begin{equation}}
\newcommand{\ee}{\end{equation}}
\newcommand{\Be}{\begin{eqnarray}}
\newcommand{\Ee}{\end{eqnarray}}
\renewcommand{\thefootnote}{\fnsymbol{footnote}}
\def\cal{calix[4]arene }
\def\cals{calix[4]arenes }
\parindent 1cm

\author[]{Marco Oestereich}
\author[]{J\"urgen Gauss}
\author[]{Gregor Diezemann}
\affil[]{Department Chemie, Johannes Gutenberg-Universit\"at Mainz, Duesbergweg 10-14, 55128 Mainz, Germany}
\title{Force probe simulations using an adaptive resolution scheme}
\maketitle
\begin{abstract}
\noindent
Molecular simulations of the forced unfolding and refolding of biomolecules or molecular complexes allow to gain important kinetic, structural and thermodynamic information about the folding process and the underlying energy landscape.
In force probe molecular dynamics (FPMD) simulations, one pulls one end of the molecule with a constant velocity in order to induce the relevant conformational transitions.
Since the extended configuration of the system has to fit into the simulation box together with the solvent such simulations are very time consuming.
Here, we apply a hybrid scheme in which the solute is treated with atomistic resolution and the solvent molecules far away from the solute are described in a coarse-grained manner.
We use the adaptive resolution scheme (AdResS) that has very successfully been applied to various examples of equilibrium simulations.
We perform FPMD simulations using AdResS on a well studied system, a dimer formed from mechanically interlocked calixarene capsules.
The results of the multiscale simulations are compared to all-atom simulations of the identical system and we observe that the size of the region in which atomistic resolution is required depends on the pulling velocity, i.e. the particular non-equilibrium situation.
For large pulling velocities a larger all atom region is required.
Our results show that multiscale simulations can be applied also in the strong non-equilibrium situations that the system experiences in FPMD simulations.
\end{abstract}
\newpage
\section*{I. Introduction}
Force spectroscopy is a standard experimental technique to investigate unfolding pathways, details of the energy landscape and the mechanical properties of single biomolecules and molecular complexes\cite{Evans:2001,Kumar:2010,Woodside:2014}.
Usually, one end of the molecule is fixed in space and an external force is applied to the other end.
This force is either constant (force clamp, FC) or changes linearly in time (force ramp, FR).
In the first case one observes the extension of the system as a function of time while in the FR protocol the force measured at the pulling device is recorded as a function of the extension\cite{Zoldak:2013}.
The information extracted from these types of experiments not only allows to determine (un)folding rates, but also 
important properties of the folding landscape like the position of transition states, transition path times or the existence of stable intermediates\cite{Dudko:2015,Chung:2018}.

Molecular dynamics (MD) simulations are routinely used to investigate conformational transitions in soft matter systems and in particular the folding and unfolding of biomolecules like peptides, proteins or RNA\cite{Bottaro:2018,Georgoulia:2019}.
The mechanical properties of biological systems can be studied with atomistic resolution using the techniques of 
FPMD simulations (also called steered MD simulations)\cite{Isralewitz:2001,Sotomayor:2007}.
In most cases, however, there is a gap of up to five orders of magnitude in the time scales of such simulations and experimental realizations of force spectroscopy\cite{Franz:2020}.
Only recently it has become possible to match the time scales of FPMD simulations and experiments performed by using a high-speed atomic force microscope to study the unbinding of a streptavidin - biotin complex\cite{Rico:2019}.
In general, however, it is challenging to reach the experimentally relevant long time scales using atomistic FPMD simulations.
One reason lies in the need of a rather large simulation box and the resulting large number of solvent molecules.
Additionally, a large number of (un)folding trajectories are required to allow for a meaningful statistical analysis of the results.
In order to speed up FPMD simulations, some of the well established techniques of coarse graining have successfully been applied 
to study the mechanical folding pathways of proteins and of RNA\cite{Hyeon:2005,Best:2008}.
However, by using these techniques that employ simplified interaction potentials and reduced numbers of particles the details of the formation and rupture of noncovalent bonds cannot be studied with atomistic resolution.
Markov State Models (MSMs) allow to study the kinetics of conformational transitions on long time scales using dynamical information from short atomistic simulation runs\cite{Husic:2018,Noe:2019} and they have successfully been applied to extent the dynamical range of FPMD simulations\cite{Ghosh:2017,G89}.
Also methods that are developed directly to increase the efficiency of FPMD simulations are available\cite{Ozer:2010,Booth:2016}.

In most cases the primary interest of FPMD simulations lies in the study of the mechanical (un)folding kinetics of the solute and  the dynamics of the solvent molecules only plays a minor role.
Therefore, mixed resolution schemes that treat the solute in an all-atom (AA) manner and the solvent in a coarse-grained (CC) way should be applicable.
There are different methods to set up mixed resolution schemes. 
One hybrid method that is particularly well suited to treat systems in which no exchange between particles treated with different resolution takes place uses the definition of virtual interaction sites\cite{Rzepiela:2011}.
These virtual sites are positioned at the center of mass of a group of atoms in the AA part of the system.
The CG forces acting on the virtual sites are then distributed uniformly among the neighboring atoms to achieve the coupling between the different parts of the system.
We have applied this methodology to the special non-equilibrium situation encountered in FPMD simulations and we have found that the scheme is applicable in principle but the accuracy is not comparable to the one in equilibrium situations\cite{G88}.
Other approaches to enable simulations with mixed resolution are (among others) the method developed by Izvekov and Voth\cite{Izvekov:2009} and the adaptive resolution scheme (AdResS) developed by Kremer, Delle Site and others\cite{Praprotnik:2006,Krekeler:2018}.  
In the present paper, we use the latter methodology for a CG description of the solvent allowing the solute to be described in an AA manner in FPMD simulations.
The AdResS is based on a partitioning of the simulation box into a region with AA resolution, one with CG resolution and a crossover or hybrid region which allows for particle exchange between the regions of different resolution.
Thus, in our study, the AA region consists of the solute and some surrounding solvent molecules while the solvent that is not in the immediate neighborhood of the solute is treated in a CG manner.
As in FPMD simulations the system is strongly driven out of thermal equilibrium, it is not clear a priori that methods developed for equilibrium simulations are also applicable in these situations.

As in earlier investigations, we will perform FPMD simulations using a \cal catenane dimer in mesitylene solvent as a  model system.
Apart from the mentioned hybrid simulations, we have investigated this system in AA simulations and have found that its reversible unfolding kinetics can well be understood in terms of a simple two-state model\cite{G69,G74,G77}.
In equilibrium, the two \cal "cups" form a complex stabilized by a ring of 16 hydrogen bonds (H-bonds).
Complete dissociation of the cups is prevented by a set of four intertwined aliphatic loops consisting of 14 methylene groups each.
We will use mesitylene as a solvent because due to its aprotic nature it does not form H-bonds that interfere with the intramolecular H-bonds between the two \cal monomers.

The paper is organized as follows.
In the next Section, the computational details are presented including a brief recapitulation of the AdResS methodology and the presentation of results of equilibrium AdResS simulations of the \cal dimer system. 
We then compare the results of FPMD simulations performed employing AdResS to those of AA simulations  and close with some concluding remarks.
\section*{II. Computational Methodology}
\subsection*{1. All-atom simulations}
\begin{figure}[h!]
\centering
\includegraphics[width=6.7cm]{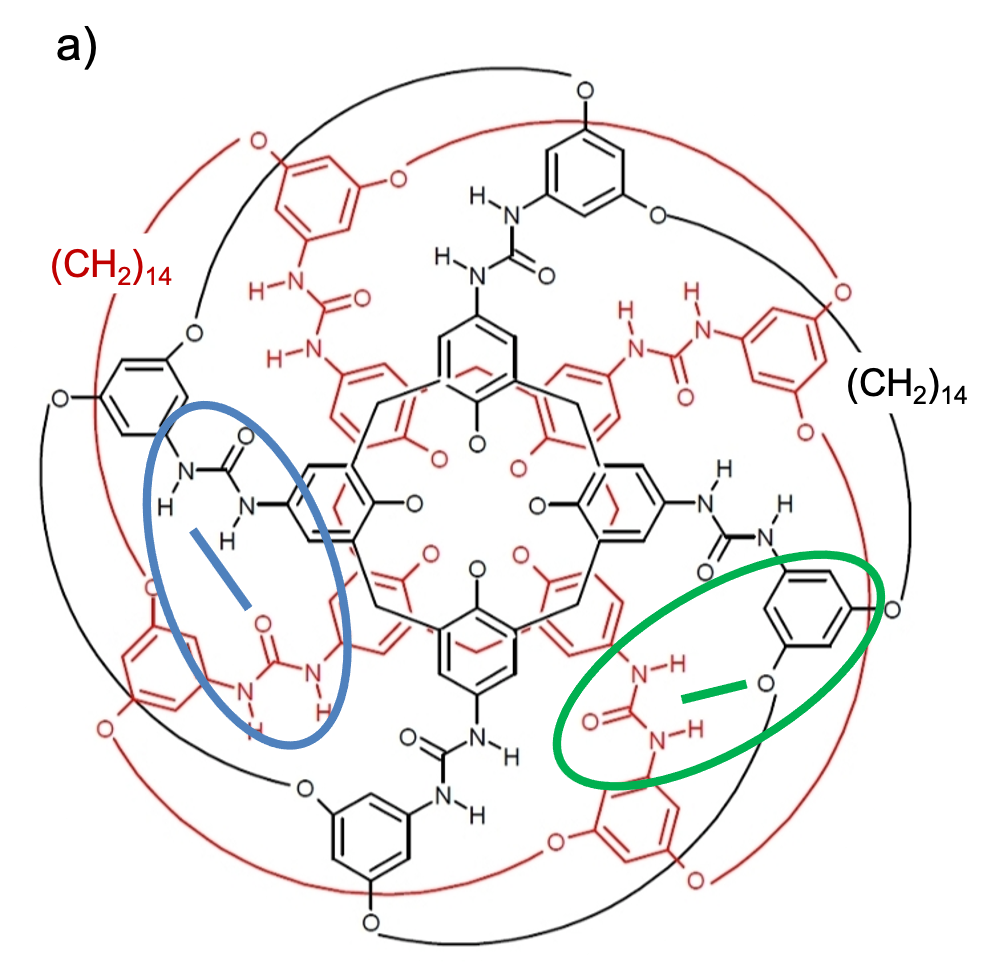}
\hspace{0.4cm}
\includegraphics[width=6.3cm]{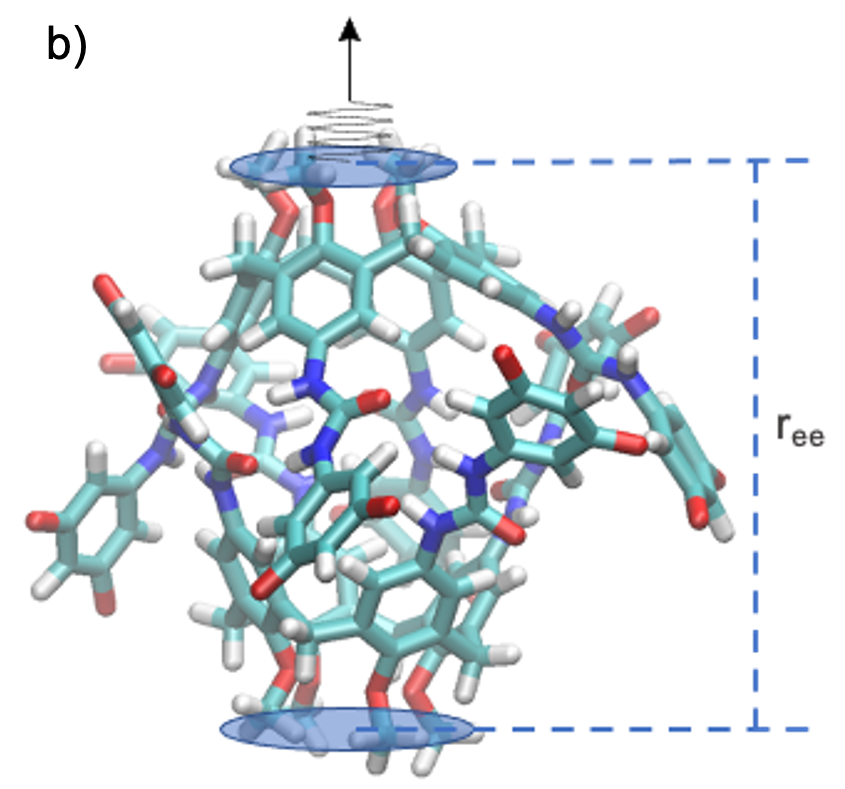}
\caption{{\bf a)} Chemical structure of the \cal catenane dimer with the H-bonds indicated. Blue: Urea-Urea (UU) bonds stabilizing the closed structure; Green: Urea-Ether (UE) bonds stabilizing the open structure.
The methyl groups at the narrow rim of the calixarene cups are omitted.
{\bf b)} Stick model of the \cal dimer along with the definition of the reference group and the pulled group used in the setup of the FPMD simulations.
These groups are defined as the center of mass of the methoxy-carbon atoms at the narrow rim of one \cal monomer.
The reference group is fixed in space and the pulled group is moved along the vector connecting the two goups using a harmonic potential. The end-to-end distance $r_{\rm ee}$ is defined as the distance between the two groups.
The aliphatic loops are omitted for clarity.
}
\label{Plot1}
\end{figure}

All AA simulations were performed using the GROMACS 2018.4 program package employing the OPLS-AA force field\cite{Hess:2008,opls2,opls1}.
We used a stochastic dynamics integrator \cite{Goga:2012} at a temperature of 298 K with a friction constant of 0.1 ps.
All bonds were constrained using the LINCS algorithm\cite{Hess:1997} allowing for a time step of 2 fs.
Short range electrostatic and van der Waals interactions were computed using a cut-off of 1.2 nm.
The long range Coulomb interactions was treated using the reaction field method with a relative dielectric constant of 2.4.
For the van der Waals interactions, we applied a dispersion correction\cite{Allen:1987}. 
For the AA simulations the neighbor list was updated after 25 simulation steps and for the CG simulations a pair list with a cutoff of 1.37 nm was used.
We used Cartesian periodic boundary conditions in all simulations.
We performed an energy minimization starting with a (7.5 nm)$^3$ cubic simulation box containing one \cal catenane dimer and 1780 mesitylene molecules, cf. Fig.\ref{Plot1}.
After this, the system was equilibrated in the canonical ensemble for 200 ps.
Then it was coupled to a Berendsen barostat \cite{Berendsen:1984} with a time constant of 0.5 ps and an isothermal compressibility of $\rm 8.26\cdot10^{-5}$ bar$^{-1}$.
The box size was determined to be (7.49 nm)$^3$ for a pressure of 1 bar and this was used in all AA simulations.
All production runs were performed in the canonical ensemble using this box size.
We mention that this box size is larger than the box size we used in earlier investigations ((5.8 nm)$^3$)\cite{G86, G96}. 
However, the larger box size used here is also used for the AdResS simulations and therefore a direct comparison is possible.
\subsection*{2. Coarse-grained potentials}
\begin{figure}[h!]
\centering
\vspace{-0.25cm}
\includegraphics[width=8.0cm]{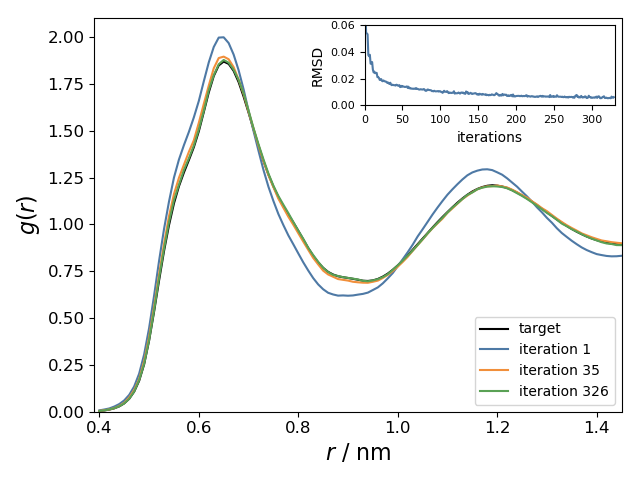}
\vspace{-0.5cm}
\caption{
RDF for the center of mass of the mesitylene molecules for different steps in the IBI procedure of the calculation of the CG pair potential.
The inset shows the root mean square deviation of the RDF $g(r)_i$ relative to the reference RDF $g(r)_{ref}$, cf. eq.(\ref{PMF.CG})
}
\label{Plot2}
\end{figure}

The CG potentials for the solvent molecules were computed using the iterative Boltzmann inversion (IBI) method\cite{Reith:2003}.
IBI relates the free-energy of a pair of particles to the logarithm of the radial distribution function (RDF) to obtain a potential of mean force (PMF) as a function of the distance $r$ between the particles\cite{Reith:2003,Hanke:2017}.
Starting from the expression for the PMF, $U(r)=-k_BT\ln(g(r))$, with $g(r)$ denoting the RDF and $k_B$ the Boltzmann constant, one obtains the effective pair potential iteratively.
The initial PMF is estimated from the reference RDF, $U(r)^{CG}_0=-k_BT\ln(g(r)_{ref})$ and the iteration cycle is defined by
\be\label{PMF.CG}
U(r)^{CG}_{i+1}=U(r)^{CG}_i+k_BT\ln\left({g(r)_i\over g(r)_{ref}}\right)
\ee
The mesitylene molecules were treated as spheres and the RDF of the center of mass was used in the determination of the PMF.
In the iterative scheme we used a cut-off for the potential of 1.2 nm and each simulation run had a duration of 200 ps (the first 20 ps were omitted for equilibration).
The effective pair potential was smoothed using cubic splines and we iterated the procedure 325 times without pressure correction.
The simulations were performed using the program package VOTCA\cite{Ruhle:2009, Mashayak:2015} (VOTCA 1.3.)
and the reference RDF was obtained from an AA simulation (25 ns) of a pure mesitylene system in a box with size (7.49 nm)$^3$.
The results for the RDF are presented in Fig.\ref{Plot2}. 
Without showing the results here, we note that the resulting PMF is of a purely repulsive nature and can approximately be described by an inverse power law potential with an exponent in the range of 12, cf. the Supporting Information.

It is well known that coarse graining speeds up the dynamics relative to AA simulations\cite{Fritz:2011,Izvekov:2006}. 
We quantified this effect by a measurement of the diffusion coefficients which are given by 
$D_{\rm CG}=11.8\cdot10^{-10}$m$^2$s$^{-1}$ and $D_{\rm AA}=1.7\cdot10^{-10}$m$^2$s$^{-1}$.
Note that these values differ from the corresponding ones given in ref.\cite{G88} 
($D_{\rm CG}=8.8\cdot10^{-10}$m$^2$s$^{-1}$ and $D_{\rm AA}=7.2\cdot10^{-10}$m$^2$s$^{-1}$). 
We attribute this to the different box sizes, different treatment of the long range electrostatic interactions and different integrators.
\subsection*{3. Adaptive resolution scheme}
\begin{figure}[h!]
\centering
\vspace{-0.25cm}
\includegraphics[width=8.0cm]{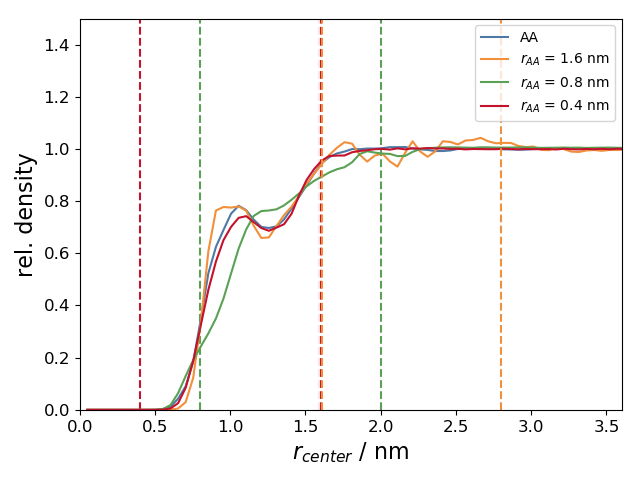}
\vspace{-0.5cm}
\caption{
Relative density of the mesitylene solvent as a function of the distance from the center, $r_{\rm center}$, of the simulation box.
The vertical lines represent the boundaries of the AA region and the hybrid region.
}
\label{Plot3}
\end{figure}

For the AdResS methodology it is important to balance the chemical potential in the AA region and the CG region. 
This is accomplished by the computation of the thermodynamic force that is calculated iteratively according to
\be
{\bf F}^{TD}_{i+1}({\bf x}) = {\bf F}^{TD}_{i}({\bf x}) - {M\over\kappa_T\rho_{ref}^2} {\bf\nabla} \rho_i({\bf x})
\ee
and works on the CG part in the hybrid region allowing the exchange of molecules between the different regions\cite{Krekeler:2018,Fritsch:2012}.
Here, $M$ is the mass of the molecules, 
$\kappa_T$ is a constant conceptionally related to the isothermal compressibility 
and $\rho_i({\bf x})$ the density in the $i$th iteration step.
The iteration starts with the initial density and a vanishing thermodynamic force.
We have used a spherical setting with different radii of the AA region and a constant slab thickness of the hybrid region of $s_{\rm Hy}=1.2$ nm.
The range of the thermodynamic force was enlarged by 0.2 nm in order to incorporate molecules that are only partly placed in the hybrid region.
We varied the radius of the AA region, $r_{\rm AA}$, in order to study the dependence of the results on this choice.
The values used are $r_{\rm AA}=1.6$, $0.8$ and $0.4$ nm. 
Since the average end-to-end distance in equilibrium is about $1.4$ nm, this means that for the smallest value of 
$r_{\rm AA}$ parts of the \cal dimer are located in the hybrid region. 
The center of the \cal dimer was always kept in the center of the simulation box ensuring that the distance between the pulled group and the reference group to the border of the CG region stayed the same.
We used a force capping methodology in the hybrid region \cite{Zavadlav:2014} with a maximum force of 
$F_{max}=2.5\cdot10^{5}$ kJ/(molK).
This ensures that the forces between two particles at the border between the CG region and the hybrid region do not increase too much. 
Force capping is applied because the CG potential depends only on the distance of the center of mass of the mesitylene molecules and is independent of the relative orientation of the molecules.
Therefore, the distance between parts of the molecules can become very small.
All simulations had a duration of 1 ns with the first 400 ps omitted due to equilibration. 
We performed 88 iterations for $r_{\rm AA}=1.6$ nm, 149 for $r_{\rm AA}=0.8$ nm and 244 for $r_{\rm AA}=0.4$ nm in order to obtain flat density profiles throughout the simulation box.
(For the smaller AA regions a flat density profile in the hybrid region is more important.)
In Fig.\ref{Plot3}, we plot the density as a function of the distance from the center of the simulation box which coincides with the center of mass of the \cal dimer.
Since the \cal dimer resides in the center of the box, the solvent density vanishes for small values of 
$r_{\rm center}$.
For larger distances, the density follows the one of the AA simulation to a very good approximation.
Due to the structure of the \cal dimer the distribution of the solvent density in the immediate neighborhood of the solute is not isotropic. This is the reason for the appearence of the hump like structure for $r_{\rm center}\sim1$ nm. 
For some further information regarding the thermodynamic force we refer to the Supporting Information.
\subsection*{4. Equilibrium simulations}
\begin{figure}[h!]
\centering
\includegraphics[width=7.5cm]{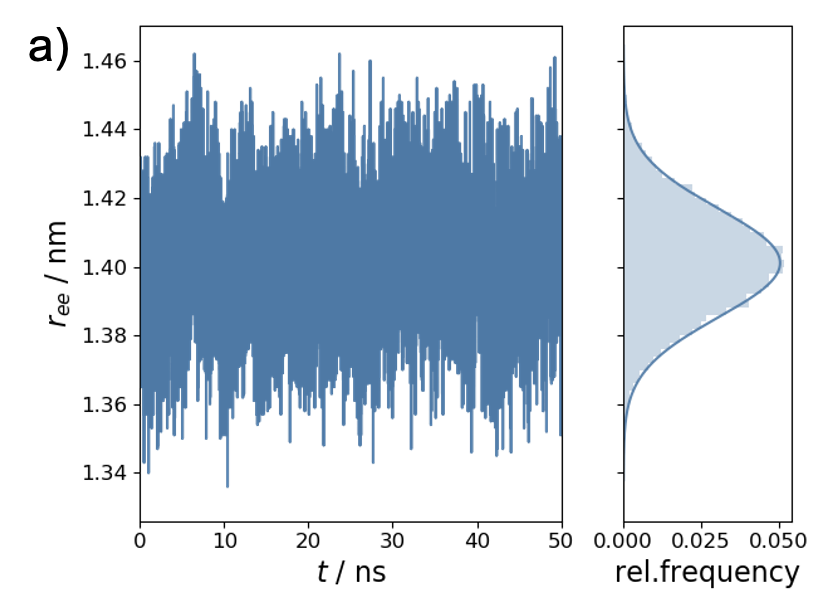}
\hspace{0.4cm}
\includegraphics[width=7.5cm]{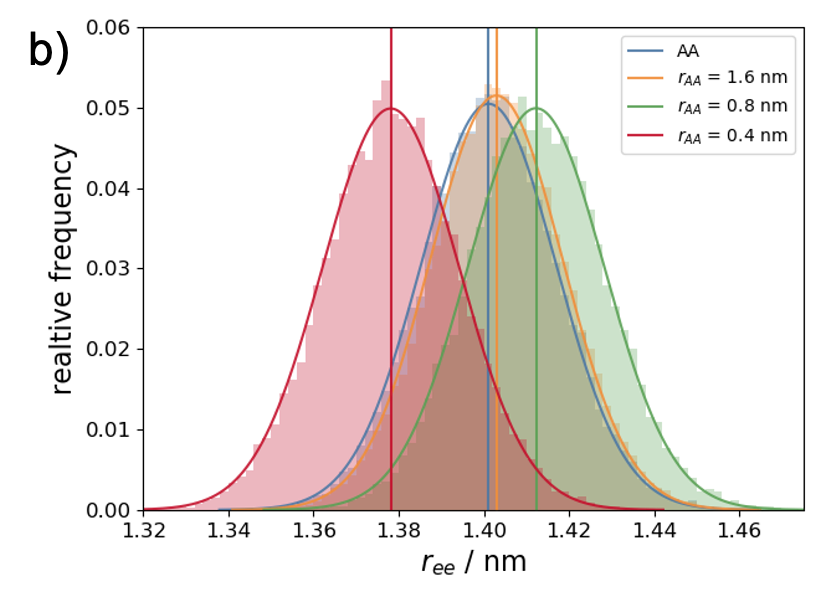}
\vspace{-0.5cm}
\caption{{\bf a)} Left: Time evolution of the end-to-end distance from an AA simulation; Right: Probability distribution of the observed values of $r_{\rm ee}$.
{\bf b)} Probability distributions for different radii of the AA region in AdResS simulations.
The full lines represent fits to Gaussians.
}
\label{Plot4}
\end{figure}

The AdResS is well known to reproduce the AA results in equilibrium simulations for a number of different situations\cite{Krekeler:2018}. 
In order to assure that it also works for our particular system, we performed equilibrium simulations and monitored the most important structural features. 
In Fig.\ref{Plot4} (a) we show the end-to-end distance $r_{\rm ee}$ as a function of the simulation time and its probability distribution for a 50 ns AA simulation.
We additionally present the distributions for AdResS simulations using various values for the radius of the AA region, $r_{\rm AA}$, in Fig.\ref{Plot4} (b).
It can be seen that the average value of $r_{\rm ee}$ is almost independent of the size of the AA region.
Only for the smallest value of $r_{\rm AA}=0.4$ nm, it differs from the AA value by about 1.5\%.
From the similarity of the widths of the distributions we conclude that also the fluctuations are very well sampled by the AdResS simulations using the given parameters.
We have also monitored the number of UU-bonds stabilizing the closed structure. 
In all simulations one observes that most of the time (more than 50\%) the maximum number of 16 UU-bonds are formed and there are quite frequent fluctuations in which one or two bonds open.
However, there are hardly any significant differences between the AA simulation and the AdResS simulations.
We thus conclude that the AdResS simulations give a good representation of the AA results.
\section*{III.  FPMD simulations}
\subsection*{1. Simulation setup}
All FPMD simulations presented in this work were performed using the FR protocoll.
We used two modes, a pull mode and a relax mode where after pulling the dimer into the open conformation the pulling direction is inverted and all other parameters remain the same.
The reference group was fixed in space and a harmonic potential was applied to the pulled group, where the groups are defined as in Fig.\ref{Plot1}.
The force measured at the spring is given by
\be\label{F.def}
F = K(V\cdot t - z(t))
\ee
Here, K is the spring constant, V the pulling velocity, and $z(t)$ denotes the deviation of the position of the pulled group from its initial value. 
Note that the extension is defined as $x=V\cdot t$ and the so-called loading rate is given by the product of the force constant and the pulling velocity, $\mu=K\cdot V$.

In FPMD simulations the system is driven out of equilibrium and it is pulled through the solvent.
In our earlier simulations employing virtual sites\cite{G88}, the coarse graining of the solvent did not have a dramatic impact on the equilibrium properties of the system, but the rupture forces were significantly reduced relative to those obtained from AA simulations.
Therefore, in the present study, we investigate the dependence of the results of AdResS simulations on the size of the AA region, the pulling velocity and the stiffness of the pulling device.

As mentioned above, we used a box of size of (7.49 nm)$^3$ for all AA simulations and the AdResS simulations although we have found in earlier studies of the \cal catenane dimer system that a box length of 5.8 nm is sufficient for all FPMD simulations performed so far\cite{G86, G96}.
A box of this size appears to be a good compromise between the two extreme scenarios that have to be considered in the case of FPMD simulations.
For quasistatic pulling, $r_{\rm ee}$ approximately follows the pulling protocol $r_{\rm ee}\simeq x_{\rm max}$ while for fast pulling one expects $r_{\rm ee}\ll x_{\rm max}$ where $x_{\rm max}$ denotes the extension reached at the end of the pulling simulation.
For a typical value of $x_{\rm max}\sim4$ nm a box of size (5.8 nm)$^3$ is large enough.
Such a box contains one dimer and 700 mesitylene molecules with 21 atoms per molecule.
In order to estimate the relevant number of atoms in an AdResS simulation, we compute the average number of mesitylene molecules in the respective regions by assuming a homogeneous density for simplicity.
For instance, using $r_{\rm AA}=1.6$ nm, we have about 90 mesitylene molecules in the AA region, 380 in the hybrid region and 1315 in the CG region.
As a result of this naive estimate, we have to treat explicitly only approximately half of the number of solvent particles as compared to the AA simulations.
Using noncubic boxes and nonspherical AA regions in the AdResS simulations this ratio can even further be reduced.
Furthermore, adaptive schemes to determine the size of the AA region are also expected to be very effective.
However, a direct estimate of the computational efficiency of the AdResS simulations as compared to the AA FPMD simulations is not possible with the present preliminary implementation of AdResS.

\begin{figure}[h!]
\centering
\vspace{-0.25cm}
\includegraphics[width=9.0cm]{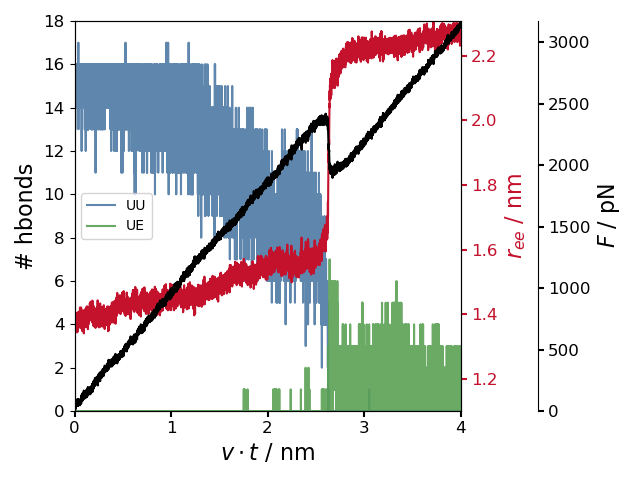}
\vspace{-0.5cm}
\caption{End-to-end distance (red), force (black), number of UU-bonds (blue) and UE-bonds (green) as a function of the extension $V\cdot t$ for a representative AA FPMD simulation in the pull mode.
The parameters are $K=1$ N/m and $V=1$ m/s ($\mu=1N/s$).
}
\label{Plot5}
\end{figure}

For a brief presentation of the results of AA simulations, in Fig.\ref{Plot5} we show examples of the most important observables as a function of the extension,
$x=V\cdot t$, for a pulling simulation with $K=1$ N/m and $V=1$ m/s (i.e. $\mu=1$ N/s).
Shown are the end-to-end distance $r_{\rm ee}$ (red), the force measured at the spring attached to the pulled group, $F$,  (black) and additionally the number of UU-bonds (blue) and UE-bonds (green). 
The transition from the closed state to the open state that takes place at an extension of roughly 2.6 nm is observable in all quantities.
The end-to-end distance increases almost linearly from the equilibrium value of 1.4 nm to about 1.55 nm and jumps at the transition to 2.1 nm. 
The force also increases linearly until there is a rip in the force versus extension curve (FEC) after which it  increases again.
If the molecular energy landscape is assumed to be harmonic, one can extract the corresponding stiffness from the slope of the FECs\cite{G67,G77,G86}.
It is also evident that the number of UU-bonds (cf. Fig.\ref{Plot1}) slowly decays for small extensions and then abruptly drops to zero at the transition point.
At this point, the UE-bonds stabilizing the open state are formed.
Due to the non-equilibrium nature of the pulling procedure the maximum number of 8 UE-bonds is not reached.
We mention that the behavior in the relax mode simulations is quite similar for the chosen parameters albeit with a finite hysteresis.
For very fast pulling the transition becomes irreversible in the sense that in the relax mode the closed state is no longer reached\cite{G74}.
\subsection*{2. Force versus extension curves}
\begin{figure}[h!]
\centering
\vspace{-0.25cm}
\includegraphics[width=9.0cm]{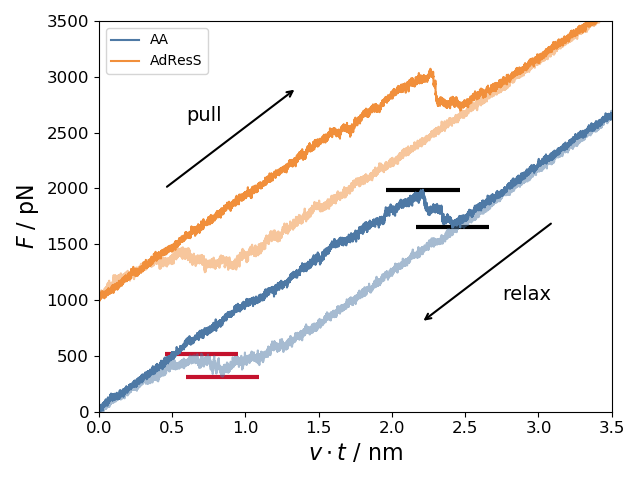}
\vspace{-0.5cm}
\caption{Examples of FECs for AA simulations and AdResS simulations for $K=1$ N/m and $V=1$ m/s.
The results for the AdResS simulations (with $r_{\rm AA}=1.6$ nm) are shifted by 1000 pN for better visibility.
The horizontal bars indicate the characteristic forces $F^{\rm min}_{\rm rupt}$, $F^{\rm max}_{\rm rupt}$ (black) and
$F^{\rm min}_{\rm rejoin}$, $F^{\rm max}_{\rm rejoin}$ (red).
}
\label{Plot6}
\end{figure}

In Fig.\ref{Plot6} we show examples of FECs as obtained from AA simulations and from AdResS simulations with 
$r_{\rm AA}=1.6$ nm for $K=1$ N/m and $V=1$ m/s.
The simulations were always performed in the same way. 
First a pulling simulation was performed until the extension $V\cdot t$ reached a value of 4 nm and then a relax mode simulation followed.
Therefore, for large extensions the two curves are on top of each other.
The hysteresis demonstrating the non-equilibrium nature of the FPMD simulations is manifested by the different extensions at which the respective transitions take place.
Only for quasistatic pulling, i.e. very small pulling velocities, the results of the pull mode and the relax mode simulations coincide.
It is evident that the FECs for the AA simulations and the AdResS simulations are very similar for the parameters used and the chosen realization.

In order to test the applicability of the AdResS methodology when applied to FPMD simulations in more detail, we used two values for the loading rate, $\mu=1$ N/s and $\mu=10$ N/s. 
These are quite high values for $\mu$ and therefore the system is driven strongly out of equilibrium.
Furthermore, we used different values of the pulling parameters as given in table \ref{Table1}.
\begin{table}[h!]
\centering
\begin{tabularx}{0.425\textwidth}{c|l|l}
& $\mu=1.0$N/s & $\mu = 10.0$N/s  \\ \hline
$K$[N/m] & $V$[m/s] & $V$[m/s] \\ \hline
1.0 & 1.0   & 10.0 \\
2.0 & 0.5   & 5.0 \\
4.0 & 0.25  & 2.5 \\
8.0 & 0.125 & 1.25\\
\end{tabularx}
\caption{Combinations of spring constant $K$ and pulling velocity $V$ used in the simulations for varying loading rate $\mu=K\cdot V$.
For each set we performed 300 AA simulations and 300 AdResS simulations for $r_{\rm AA}=0.4$ nm, 0.8 nm, 1.6 nm.}
\label{Table1}
\end{table}
As the conformational transition from the closed state to the open state and vice versa are stochastic processes, the rupture forces and the rejoin forces will vary in different realizations of the FPMD simulation using the same parameters.
Therefore, in order to provide a meaningful statistical analysis of the results, we performed 300 simulations for each set of parameters.

\begin{figure}[h!]
\centering
\vspace{-0.25cm}
\includegraphics[width=9.0cm]{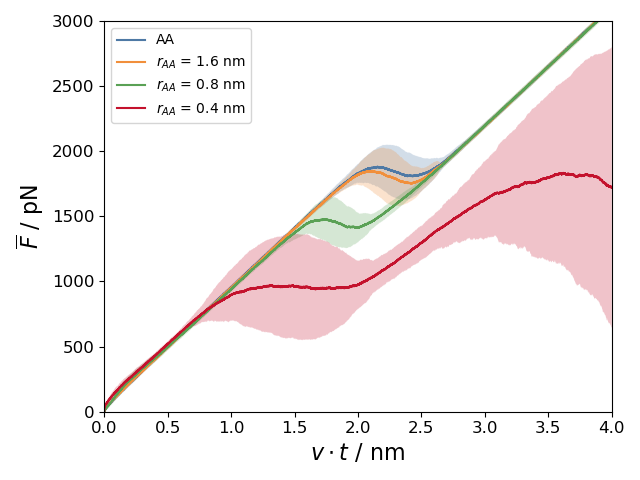}
\vspace{-0.25cm}
\caption{Averaged FECs for AA simulations and AdResS simulations using different values for $r_{\rm AA}$ as indicated.
The pulling parameters are $K=1$ N/m and $V=1$ m/s.
}
\label{Plot7}
\end{figure}

We first consider the FECs as they have been shown for one example in Fig.\ref{Plot6}.
In Fig.\ref{Plot7} we show averaged FECs\cite{G67,Seifert:2002} obtained from pull mode simulations.
The shaded areas are meant to represent the width of the distributions (the second moment).
It is evident that there is a very good agreement between the results of the AA simulations and of the AdResS simulations for $r_{\rm AA}=1.6$ nm.
For $r_{\rm AA}=0.8$ nm, the system appears to be softer in the sense that the rupture force is smaller.
The slopes of the averaged FECs, however, are the same meaning that the molecular stiffness is unaltered.
This changes for the smallest AA region used, $r_{\rm AA}=0.4$ nm. 
Here, the rupture event appears at still smaller force and has more the form of a broad shoulder than of a rip.
However, the most prominent difference to all other simulations is given by the fact that the slope in the open state is smaller than in the other cases and furthermore for large extensions, $x\gtrsim3.5$ nm the force decreases again indicating that the system becomes unstable.
This can be understood from the fact that the end-to-end distance exceeds 2.2 nm for these extensions, cf. 
Fig.\ref{Plot5}.
Thus, a substantial part of the \cal dimer enters the hybrid region and the aliphatic loops are destabilized because the relevant interactions are not considered in the AdResS protocol.
Due to this failure of AdResS for such a small value of $r_{\rm AA}$ when applied to FPMD simulations we will no longer consider simulations performed using $r_{\rm AA}=0.4$ nm.

\begin{figure}[h!]
\centering
\vspace{-0.25cm}
\includegraphics[width=12.0cm]{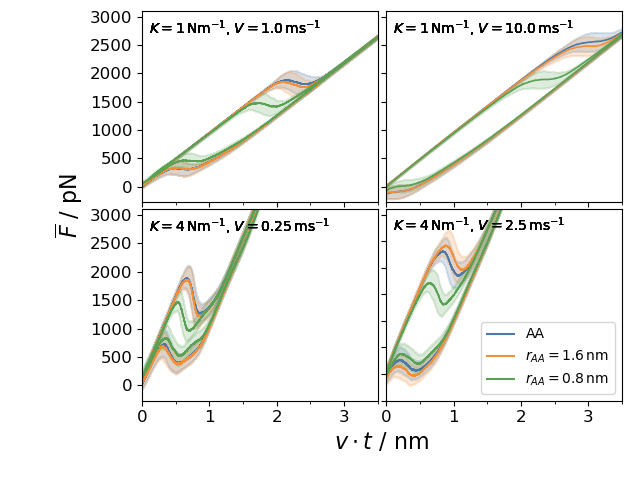}
\vspace{-0.75cm}
\caption{Averaged FECs for AA simulations and AdResS simulations for two values of $K$ as indicated and two loading rates.
Left panels: $\mu=1$ N/s; Right panels: $\mu=10$ N/s.
}
\label{Plot8}
\end{figure}

In Fig.\ref{Plot8} we show averaged FECs in the pull mode and the relax mode for various choices of the parameters used in the protocol.
The general features are well reproduced qualitatively by all AdResS simulations independent of the pulling velocity and the loading rate.
The hysteresis shows the known dependence on $V$ and also an increase of the rupture force with the loading rate is 
observed\cite{G67,G86}.
In particular, the results of the AA simulations and of the AdResS simulations for $r_{\rm AA}=1.6$ nm coincide very well for all sets of parameters in both modes.
The finite size of the AA region becomes important for $r_{\rm AA}=0.8$ nm. 
In this case, the mean rupture forces are smaller and the mean rejoin forces are somewhat larger than in case of the AA simulations.
In case of the pull mode simulations the differences of the mean rupture forces exceed the widths of the distributions.
Additionally, the differences apparently are more pronounced for the larger loading rate. 
\subsection*{3. Rupture force distributions}
In order to discuss these findings in more detail, we consider the distributions of the characteristic forces obtained from the individual FECs as shown in Fig.\ref{Plot6}.
In the present paper we will consider the mean force between the maxima and the minima in the transition region for both cases, pull and relax mode FECs,
\be\label{Frure.def}
F_{\rm rupt}={1\over2}\left( F^{\rm min}_{\rm rupt}+F^{\rm max}_{\rm rupt}\right)\,\mbox{and}\,
F_{\rm rejoin}={1\over2}\left( F^{\rm min}_{\rm rejoin}+F^{\rm max}_{\rm rejoin}\right),
\ee
cf. Fig.\ref{Plot6}.
These forces behave somewhat different from the maximum rupture force and the minimum rejoin force but capture all important features\cite{G86,G96}.
Furthermore, and more important in the present context, for reversibly binding systems like the \cal dimer investigated here, $F_{\rm rupt}$ and $F_{\rm rejoin}$ will converge to the equilibrium value, $F_{\rm eq}$, for slow pulling\cite{G67}.

\begin{figure}[h!]
\centering
\vspace{-0.5cm}
\includegraphics[width=8.0cm]{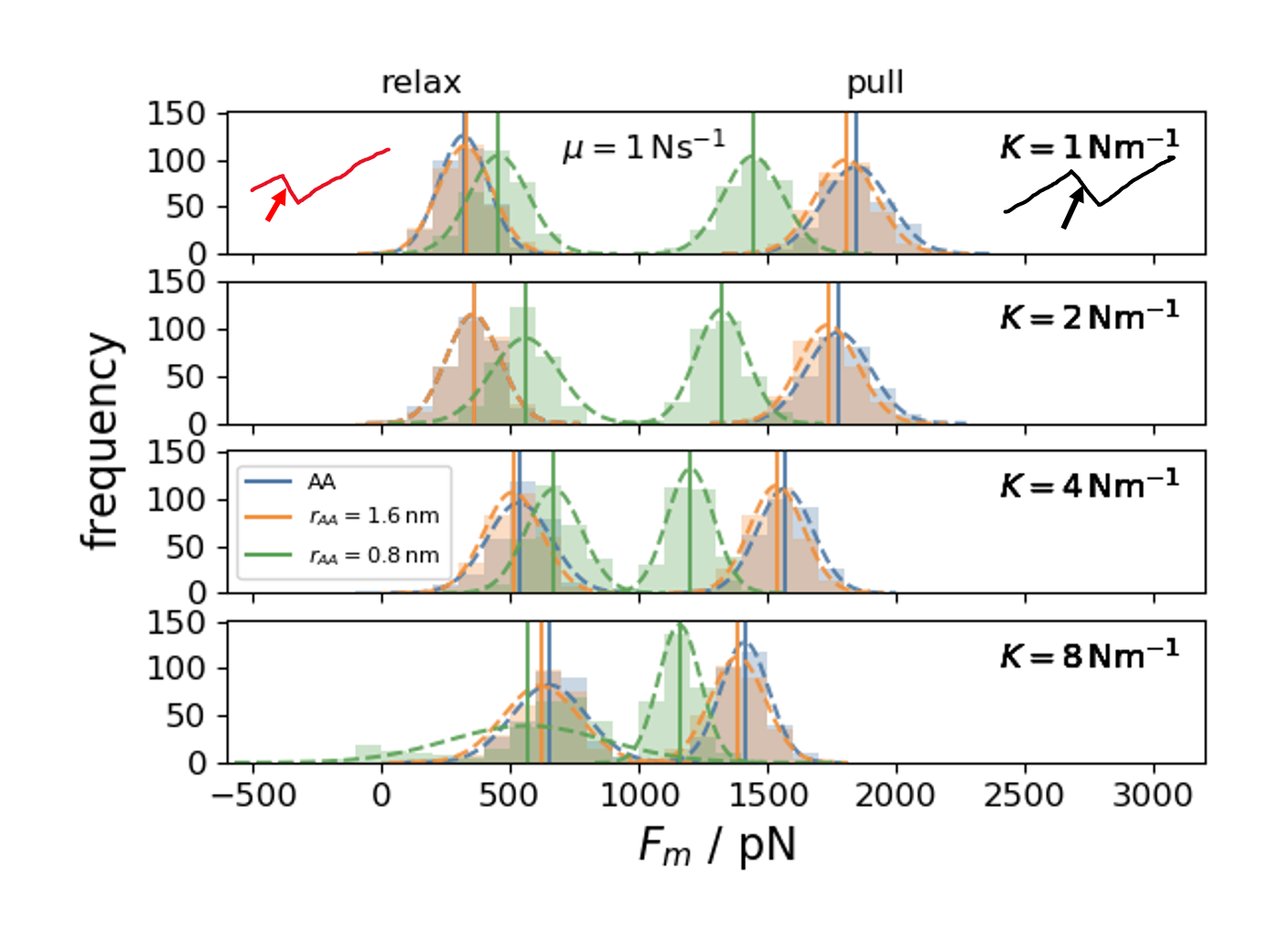}
\includegraphics[width=8.0cm]{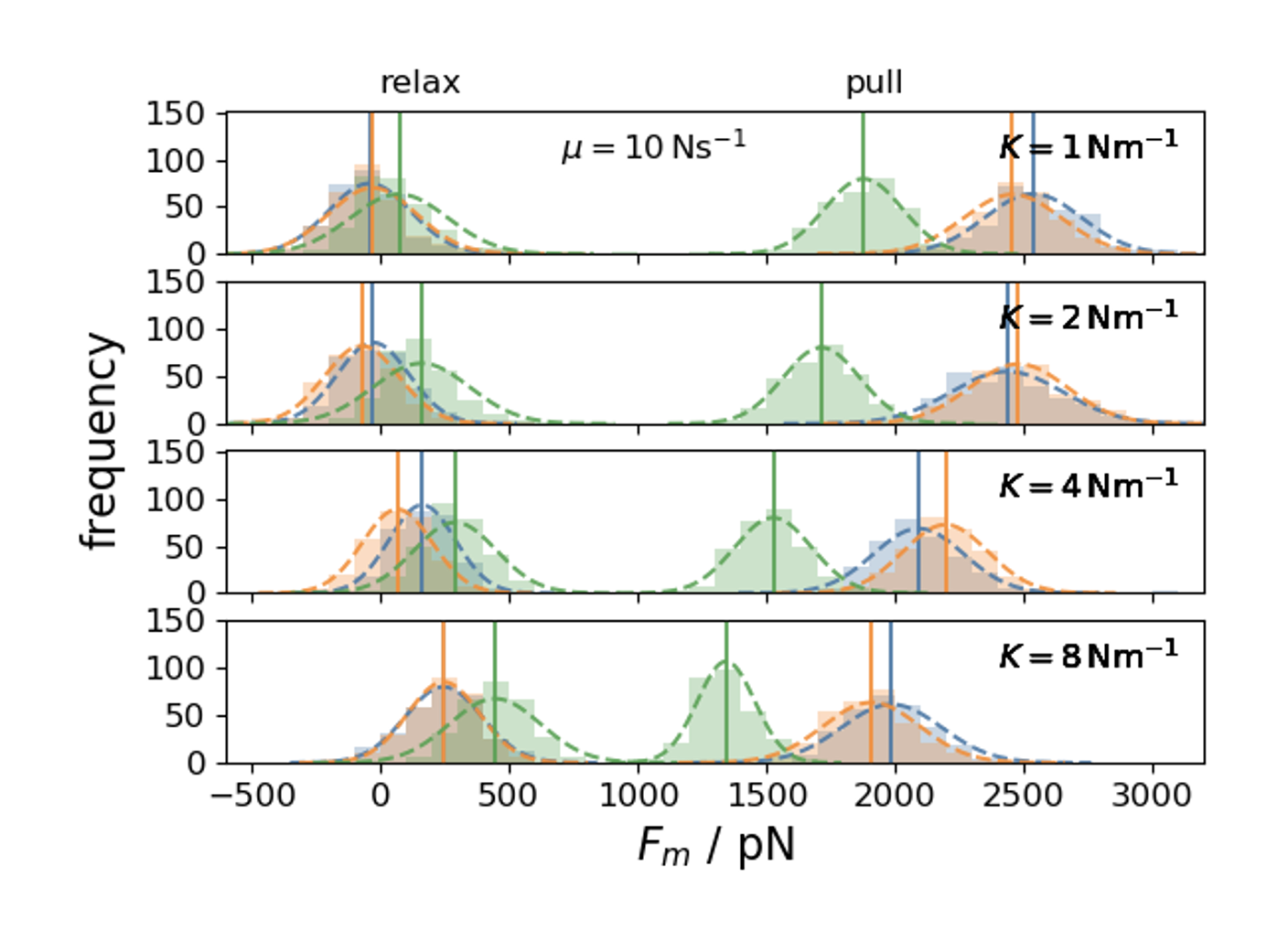}
\vspace{-1.25cm}
\caption{Distributions of rupture forces $F_{\rm rupt}$ and rejoin forces $F_{\rm rejoin}$ defined in Eq.(\ref{Frure.def}).
The vertical bars indicate the average forces and the dotted lines represent Gaussian fits to the distributions.
}
\label{Plot9}
\end{figure}

Fig.\ref{Plot9} shows these distributions for the four values of the force constants and the concomittant pulling velocities as listed in table\ref{Table1}.
The mean rupture forces (vertical  bars in Fig.\ref{Plot9}) slightly decrease as a function of the force constant and the mean rejoin forces slightly increase.
This indicates that they will merge for vanishing pulling velocity.

As pointed out above, the AdResS simulations using $r_{\rm AA}=1.6$ nm yield results almost identical to the AA simulations.
In case of the AdReS simulations with $r_{\rm AA}=0.8$ nm the distributions of rupture forces are shifted to smaller forces while the $F_{\rm rejoin}$ distributions tend to be shifted to larger forces.
This means, in these AdResS simulations the dimer behaves as if the "distance to equilibrium" would be smaller than in the AA simulations.
This finding is similar to what we have observed in temperature dependent AA simulations for increasing temperature\cite{G97}.
We mention that the number of simulations performed (300 for each set of parameters) is not large enough for a detailed discussion of  the shape and the widths of the distributions.
Furthermore, the observed differences appear to depend only weakly on the force constant and more strongly on the loading rate.
Only for the slowest pulling velocity of $V=0.125$ m/s the difference in the rupture forces is somewhat smaller than for faster pulling.
At the same time the $F_{\rm rejoin}$ distribution becomes rather broad.
We interpret this finding as resulting from the fact that for very slow pulling, in the relax mode simulation the system resides in the open state for quite a long time and $r_{\rm ee}$ decreases only slowly.
Therefore, the higher solvent mobility in the CG region (and also in the hybrid region) apparently gives rise to some further softening of the dimer.

\begin{figure}[h!]
\centering
\vspace{-0.25cm}
\includegraphics[width=8.0cm]{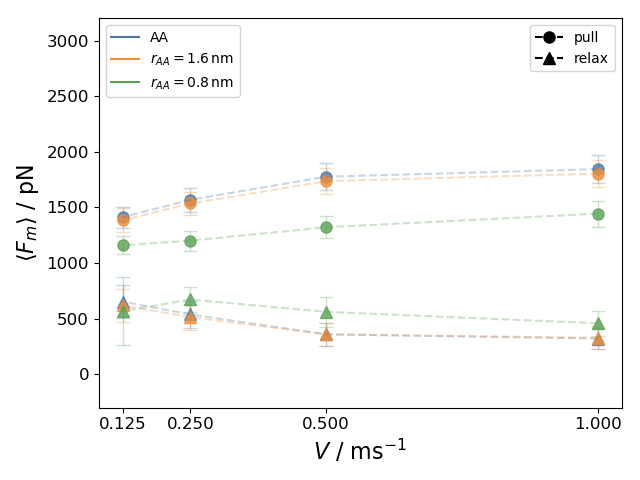}
\includegraphics[width=8.0cm]{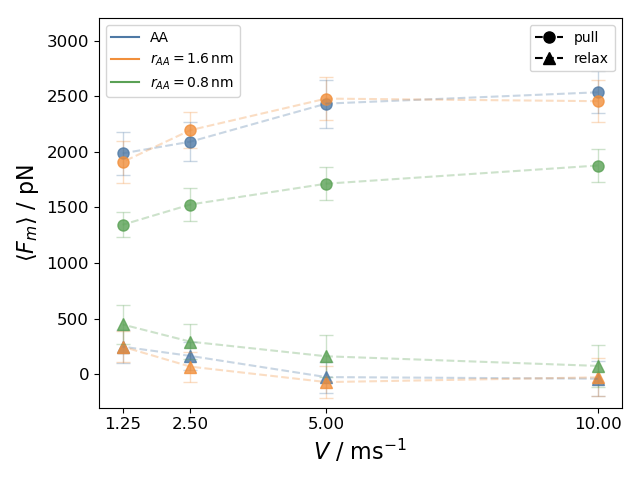}
\vspace{-1.0cm}
\caption{Mean forces $\langle F_{\rm rupt}\rangle$ and $\langle F_{\rm rejoin}\rangle$ as a function of the pulling velocity $V$.
Left: $\mu=1$ N/s; Right: $\mu=10$ N/s.
}
\label{Plot10}
\end{figure}

The most prominent features that can be observed in the force distributions are also reflected in the mean forces.
In Fig.\ref{Plot10}, we present the mean rupture and the mean rejoin forces as a function of the pulling velocity.
An increase of the rupture force and a slight decrease of the rejoin force is observed for both loading rates.
The differences of the results for $r_{\rm AA}=0.8$ nm to the other simulation results discussed above are evident and they are somewhat more pronounced for $\mu=10$ N/s than for $\mu=1$ N/s.
As mentioned above, the results regarding the differences appear to depend mainly on the loading rate.
The average difference in the values of $\langle F_{\rm rupt}\rangle$ is on the order of 400 pN for $\mu=1$ N/s and it is about 650 pN for $\mu=10$ N/s.
The differences in the mean rejoin forces is almost the same for both loading rates (approximately 150 pN).

The decrease of the rupture force for the AdResS simulations with $r_{\rm AA}=0.8$ nm indicates that the increased solvent mobility ($D_{\rm CG}/D_{\rm AA}\simeq7$) is important.
If we assume that the dynamics of the \cal dimer is dominated by activated barier crossing, the celebrated Bell model\cite{Bell:1978} yields a logarithmic dependence on the diffuson coefficient, $F_{\rm rupt}\sim\log{(\mu/D)}$\cite{Hummer:2003}.
However, the ratio between the rupture forces should not vary with loading rate.
The larger differences for $\mu=10$ N/s possibly are due to the fact that the regime of drift motion might be entered for fast pulling\cite{G86}.
In this regime the rupture force behaves as $F_{\rm rupt}(V\to\infty)\sim \sqrt{(\mu/D)}$\cite{Hummer:2003,Bullerjahn:2014} and therefore depends somewhat stronger on $\mu$ than in case of activated dynamics.

The situation apparently is somewhat different if the relax mode simulations are considered.
The differences in the mean rejoin forces are almost independent of $V$ and of $\mu$. 
The increased solvent mobility experienced by the dimer in the open state seems to have a smaller impact on the rejoin forces.
\subsection*{4. Analysis of the H-bond network}
\begin{figure}[h!]
\centering
\vspace{-0.25cm}
\includegraphics[width=8.0cm]{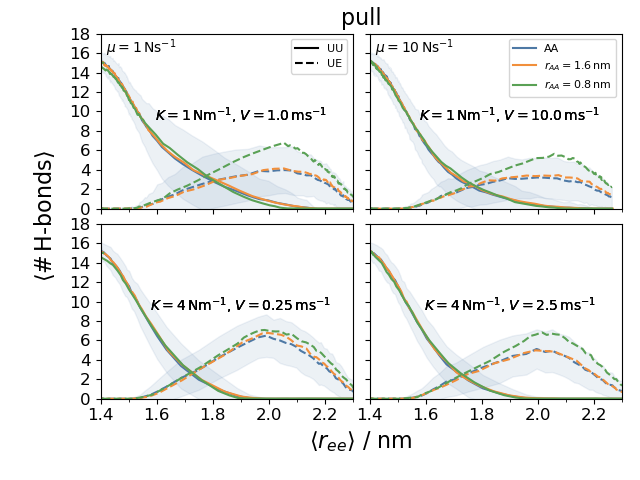}
\includegraphics[width=8.0cm]{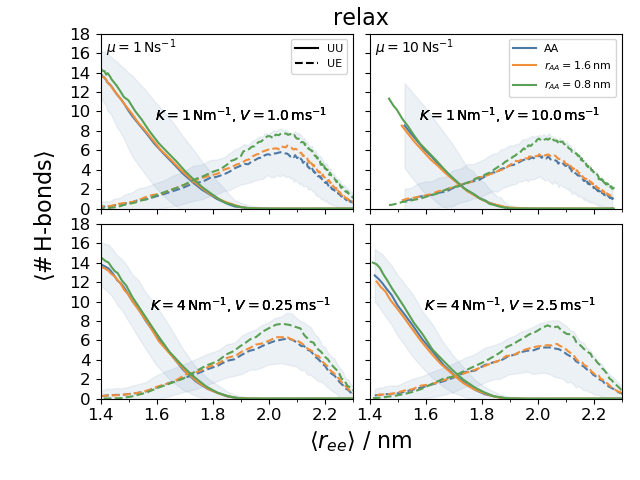}
\vspace{-0.75cm}
\caption{Averaged number of H-bonds (both UU and UE) as a function of $\langle r_{\rm ee}\rangle$.
The shaded areas indicate the width of the distributions for the AA simulations.
}
\label{Plot11}
\end{figure}

We now discuss some aspects of the dynamics of the H-bond network and some properties of the energy landscape.
As mentioned above, the network of UU-bonds stabilizes the closed state and these bonds open at the transition into the open state where the network of UE-bonds is formed, cf. Fig.\ref{Plot5}.
Since the transition is a stochastic process and takes place at different extensions depending on the pulling velocity, 
the number of H-bonds also strongly depends on the extension.
On the other hand, the different conformational states are characterized by the corresponding ranges of the 
end-to-end distance.
Therefore, in the following we consider the mean number of H-bonds as a function of the average end-to-end distance of the dimer, 
$\langle r_{\rm ee}\rangle$, and present the results in Fig.\ref{Plot11}.
This representation has been shown to depend only weakly on the pulling protocol\cite{G77,G86}.
In the pull mode, the average number of UU-bonds, $\langle \# UU\rangle$, decreases from almost 16, the maximum possible number, to zero.
The average number of UE-bonds, $\langle \# UE\rangle$, reaches a maximum at $\langle r_{\rm ee}\rangle\sim2$ nm and for further stretching they decrease again.
A similar behavior is observed in the relax mode simulations with the only difference that the maximum number of UE-bonds formed is somewhat larger than in the pull mode.
In all cases, the AdResS simulations using $r_{\rm AA}=1.6$ nm yield results that are in excellent agreement with those obtained from AA simulations. 
For simulations employing the smaller AA region, $r_{\rm AA}=0.8$ nm, $\langle \# UU\rangle$ behaves very similar to the other results while $\langle \# UE\rangle$ exceeds the values obtained from the other simulations in the region of the maximum indicating a slightly higher stability of the open state in this case as it would be expected for a somewhat smaller
loading rate.
However, these differences still are very small and within the width of the distributions.
It is interesting to note that $\langle \# UU\rangle$ is the same as for the AA simulations also in the relax mode simulations and the enhanced stability of the open state plays almost no role for the structure of the closed state.

\begin{figure}[h!]
\centering
\vspace{-0.25cm}
\includegraphics[width=10.0cm]{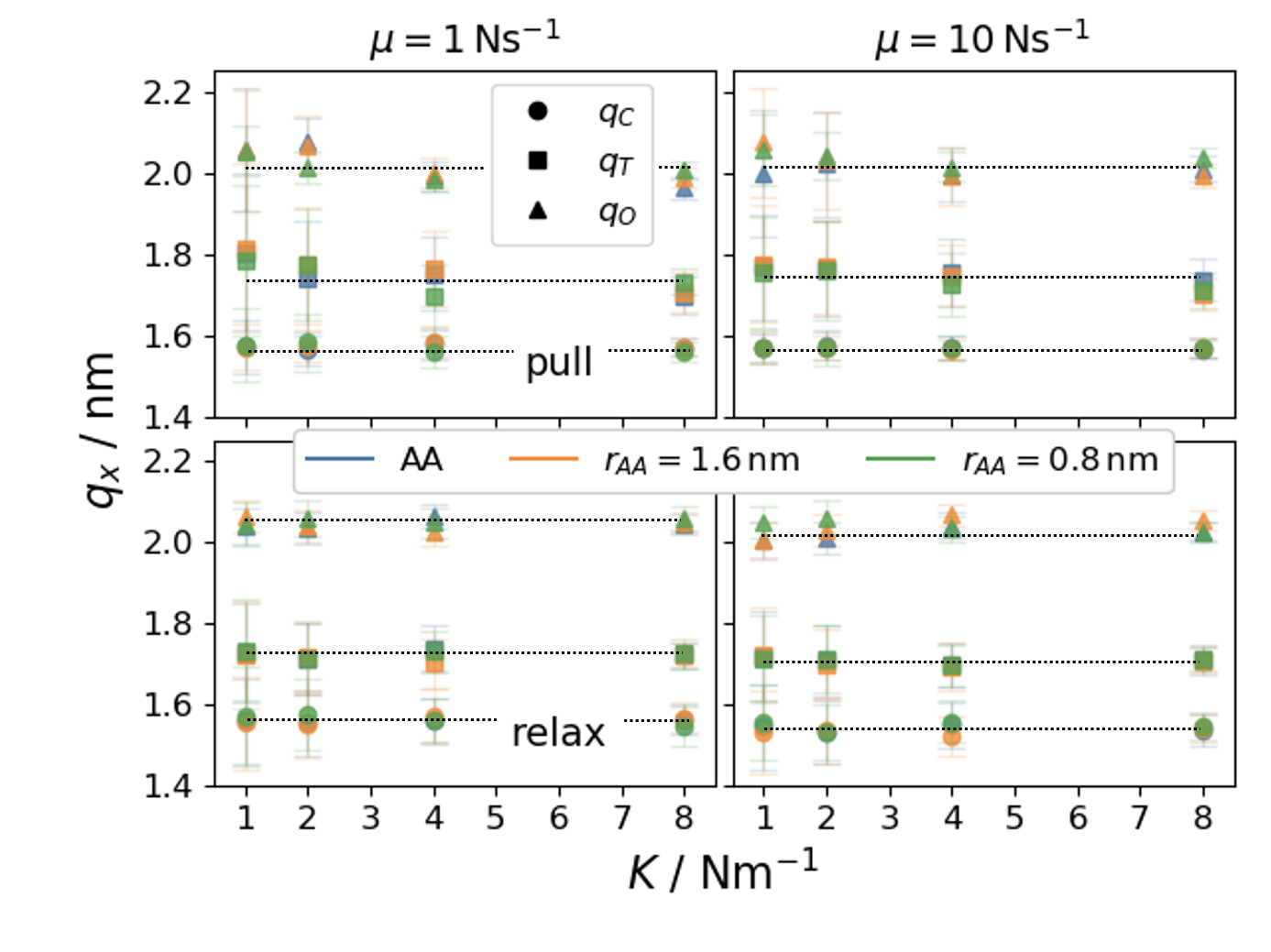}
\vspace{-0.75cm}
\caption{Locations of the minima and the transition states as defined in Eq.(\ref{qx.def}) as a function of the force constant; Left panels: $\mu=1$ N/s; Right panels: $\mu=10$ N/s.
Upper panels: pull mode; Lower panels: relax mode.
The error bars represent the standard deviations of the distributions and the dotted horizontal lines are guides to the eye.
}
\label{Plot12}
\end{figure}

As in our earlier work on the unfolding kinetics of the \cal dimer, we use the average number of H-bonds for a characterization of the energy landscape of the system.
In particular, we define the minima of the closed state ($q_C$), the open state ($q_O$) and the transition state ($q_T$) in the following way\cite{G86}:
\be\label{qx.def}
q_C=\langle r_{\rm ee}\rangle(\langle \# UU\rangle=8)
\,;\,
q_O=\langle r_{\rm ee}\rangle(\langle \# UE\rangle=\rm{max})
\,;\,
q_T=\langle r_{\rm ee}\rangle(\langle \# UU\rangle=\langle \# UE\rangle)
\ee
and present the results in Fig.\ref{Plot12}.
For both loading rates, the results are independent of the force constant and coincide within the statistical error.
We conclude from these findings that the energy landscape of the dimer is hardly altered in the AdResS simulations.
\section*{IV. Conclusions}
We have presented a detailed investigation of the applicability of the AdResS methodology to FPMD simulations.
As an example we chose a well studied system, a \cal dimer, that is known to undergo conformational transitions that can be described by a two-state model.
Additionally, the system is small enough to allow for the direct comparison with AA simulations.
We studied the system for different sizes of the AA region and a fixed slab thickness of the hybrid region.
In equilibrium, the AdResS is known to work with excellent results when compared to AA simulations and we confirmed this finding with the simulations of our system consisting of one dimer in mesitylene solvent.

Since AdResS has been developed for equilibrium simulations, it is not clear to which extent it can be applied to non-equilibrium situations.
We therefore studied the performance of AdResS as applied to FPMD simulations for high pulling velocities and large loading rates.
We found that all AdResS simulations using an AA region with a radius of $r_{\rm AA}=1.6$ nm yield results that are basically identical to the results obtained with AA simulations.
This holds for all quantities that characterize the kinetics of the conformational transition in our model system and also the structural features like the number of the H-bonds formed in each of the states.
It is important to point out that a box size comparable to the AA region in the AdResS simulations is much too small to give reliable results in AA simulations, in particular in the open state of the dimer.

The results of AdResS simulations employing a smaller AA region ($r_{\rm AA}=0.8$ nm) worked very well in equilibrium  but gave rise to deviations from the AA results in FPMD simulations.
This holds in particular when the rupture forces and rejoin forces are considered. 
Here, the system appears softer and the distance to equilibrium for a given loading rate seems to be reduced.
We attribute this to the higher solvent mobility in the hybrid region and in the CG region.
The observed differences apparently increase with increasing loading rate meaning that for larger $\mu=K\cdot V$ a larger AA region is required and the results hardly depend on the stiffness $K$ or the pulling velocity $V$ separately. 
When structural features like the number of H-bonds stabilizing the conformational states of the dimer are considered,
the AdResS simulations gave a very good representation of the results of the AA simulations.
The same holds for the characteristic features of the energy landscape as deduced from the properties of the H-bond network.

The model system considered in the present work is very simple in the sense that the (un)folding pathway is determined by two states only.
Furthermore, the solvent is of an aprotic nature and therefore does not interfere with the H-bonds stabilizing the conformations of the dimer.
However, the solvent mobility has proven to play a crucial role even in this well defined situation.
Therefore, the size of the AA region in applications of the AdResS methodology to FPMD simulations has to be chosen with care.
This holds in particular if polar solvents are considered since in that case electrostatic interactions become of major importance.
A study of this effect is under way.

In conclusion we have demonstrated that the AdResS methodology can be applied successfully to perform FPMD simulations.
The number of particles treated in an AA manner can be reduced considerably but the current preliminary implementation of AdResS for FPMD simulations does not allow for a quantitative estimate of the computational efficiency of the methodology.
\section*{Supporting material}
See the supporting information for some aspects of AdResS, the coarse grained potentials and the thermodynamic force.
\section*{Acknowledgement}
Financial support by the DFG via TRR 146 is gratefully acknowledged.
The authors gratefully acknowledge the computing time granted on the supercomputer Mogon at Johannes Gutenberg University Mainz (hpc.uni-mainz.de).
%

%\bibliography{DFS_AdResS,GregLit}
%
\end{document}